\documentclass[preprint,showpacs,preprintnumbers,amsmath,amssymb]{revtex4}

\usepackage{graphicx}
\usepackage{dcolumn}
\usepackage{bm}

\begin{document}

\preprint{Europhysics Letters}

\title{Superluminal group velocity in an anisotropic metamaterial}

\author{Hailu Luo$^{1}$}\thanks{Author to whom correspondence should be addressed.
E-mail: hailuluo@163.com}
\author{Wei Hu$^{2}$}
\author{Weixing Shu$^{1}$}
\author{Fei Li$^{1}$}
\author{Zhongzhou Ren$^{1}$}
\affiliation{$^{1}$ Department of Physics,
Nanjing University, Nanjing 210008, China\\
$^{2}$Laboratory of Light Transmission Optics, South China Normal
University, Guangzhou 510630, China}
\date{\today}

\begin{abstract}
Based on boundary condition and dispersion relation, the
superluminal group velocity in an anisotropic metamaterial (AMM)
is investigated. The superluminal propagation is induced by the
hyperbolic dispersion relation associated with the AMM. It is
shown that a modulated Gaussian beam exhibits a superluminal group
velocity which depends on the choice of incident angles and
optical axis angles. The superluminal propagation does not violate
the theory of special relativity because the group velocity is the
velocity of the peak of the localized wave packet which does not
carry information. It is proposed that a triglycine sulfate (TGS)
crystal can be designed and the superluminal group velocity can be
measured experimentally.
\end{abstract}

\maketitle

\section{Introduction}\label{Introduction}
In 1968, Veselago firstly introduced the concept of left-handed
material (LHM) in which both the permittivity $\varepsilon$ and
the permeability $\mu$  are negative~\cite{Veselago1968}. Veselago
predicted that LHM would have unique and potentially interesting
properties, such as the negative refraction index, the reversed
Doppler shift and the backward Cerenkov radiation. LHM did not
receive much attention as it only existed in a conceptual form.
After the first experimental observation using a metamaterial
composed of split ring resonators
(SRR)~\cite{Smith2000,Shelby2001}, the study of such materials has
received increasing attention over the last few
years~\cite{Kong2002,Smith2002,Ziolkowski2001,Gupta2004,Woodley2004,Lindell2001,Hu2002,Zhou2003,Smith2003,Luo2005,Smith2004a,Smith2004b}.
As noted earlier, both $\varepsilon$ and $\mu$ are necessarily
frequency dispersive in LHM. Since the frequency dispersion is
important, the superluminal propagation in the LHM takes
place~\cite{Ziolkowski2001,Gupta2004,Woodley2004}.

While negative refraction is most easily visualized in an
isotropic metamaterial, negative refraction can also be realized
in anisotropic metamaterial (AMM), which does not necessarily
require that all tensor elements of $\boldsymbol{\varepsilon}$ and
$\boldsymbol{\mu}$ have negative
values~\cite{Lindell2001,Hu2002,Zhou2003,Smith2003,Luo2005}.
Recently, Smith {\it et al.} have shown experimentally that an AMM
slab designed to provide a permeability equal to $-1$ along the
optical axis, will redirect $E$-polarized electromagnetic waves
from a nearby source to a partial
focus~\cite{Smith2004a,Smith2004b}.

In the present letter, we will investigate the superluminal group
velocity of waves in an AMM. The superluminal propagation is
induced by the hyperbolic dispersion relations associated with the
AMM.  We describe a modulated Gaussian beam incident on the
triglycine sulfate (TGS) crystal, which demonstrates in a
straightforward manner that the peak of the localized wave packet
displays interesting superluminal behavior.

\section{Hyperbolic dispersion relation}\label{sec2}
For anisotropic media, one or both of the permittivity and
permeability are second-rank tensors. We assume that the
permittivity and permeability tensors are simultaneously
diagonalizable.  If we take the optical axis as the $z$ axis, the
permittivity and permeability tensors have the following forms:
\begin{eqnarray}
\boldsymbol{\varepsilon}=\left(
\begin{array}{ccc}
\varepsilon_x  &0 &0 \\
0 & \varepsilon_y &0\\
0 &0 & \varepsilon_z
\end{array}
\right), ~~\boldsymbol{\mu}=\left(
\begin{array}{ccc}
\mu_x  &0 &0 \\
0 & \mu_y &0\\
0 &0 & \mu_z
\end{array}
\right),\label{matrix}
\end{eqnarray}
where $\varepsilon_i$ and $\mu_i$  are the relative permittivity
and permeability constants in the principal coordinate system
($i=x,y,z$). It should be noted that the real AMM constructed by
SRR is highly dispersive, both in spatial sense and frequency
sense ~\cite{Smith2004a,Smith2004b}. So these relative values are
functions of the angle frequency $\omega$. A general study on the
shape of the dispersion relation as function of the sign of these
parameters has already been offered in Ref.~\cite{Smith2003}. In
this work, we are interested in the case of AMM with hyperbolic
dispersion relation.

\begin{figure}
\includegraphics[width=10cm]{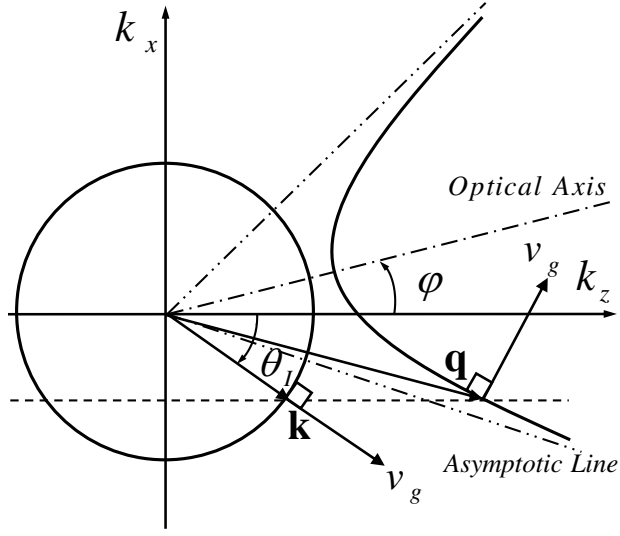}
\caption{\label{Fig1} The circle and the hyperbola represent the
dispersion relations of free space and AMM, respectively. The
incident wave vector ${\bf k}$ is parallel to the group velocity
${\bf v}_g$ in free space. Because of the anisotropy, ${\bf
v}_g=\nabla_{\bf q}\omega({\bf q})$ in the AMM is not necessarily
parallel to the refracted wave vector ${\bf q}$. $\nabla_{\bf
q}\omega({\bf q})$ must lie normal to the frequency contour,
$\omega({\bf q})=const$. }
\end{figure}

Without loss of generality, we assume that the wave vector locate
at the $x-z$  plane ($k_y=q_y=0$).  Maxwell's equations yield a
scalar wave equation for $H$-polarized field . In free space, the
accompanying dispersion relation has the familiar form:
\begin{equation}
 k_{x}^2+k_{z}^2= \frac{\omega^2}{c^2}, \label{D1}
\end{equation}
where $k_x$ and $k_z$ are the $x$ and $z$ components of the
incident wave vector, $\omega$ is the frequency, and $c$ is the
speed of light in vacuum. We assume that there is an angle
$\varphi$ between the optical axis and the $z$ axis. For the given
polarization, the waves equation yield the dispersion relation in
AMM as
\begin{equation}
\alpha q_{x}^2+\beta q_{z}^2+\gamma
q_{x}q_{z}=\frac{\omega^2}{c^2}, \label{D2}
\end{equation}
where $q_{x}$ and $q_{z}$ represent the $x$ and $z$ components of
refracted wave vector, $\alpha$, $\beta$ and $\gamma$ are given by
\begin{eqnarray}
\alpha &=&\frac{1}{ \varepsilon_x \varepsilon_z \mu_y }
(\varepsilon_x \cos^2\varphi+\varepsilon_z \sin^2\varphi),\nonumber\\
\beta &=&\frac{1}{ \varepsilon_x \varepsilon_z \mu_y }
(\varepsilon_x \sin^2\varphi+\varepsilon_z \cos^2\varphi),\nonumber\\
\gamma &=&\frac{1}{\varepsilon_x \varepsilon_z \mu_y}
(\varepsilon_z \sin 2\varphi -\varepsilon_x \sin 2\varphi).
\end{eqnarray}
We show the dispersion geometry in Fig.~\ref{Fig1}, where a plane
electromagnetic wave is incident from free space into the AMM. We
choose the $z$ axis to be normal to the interface, the $x$ axis in
the plane of the figure, and the $y$ axis out of the plane of the
figure.

We assume here that the electric field is polarized along the $y$
axis.  The $z$-component of the wave vector can be found by the
solution of Eq.~(\ref{D2}), which yields
\begin{equation}
 q_z = \frac{1}{2 \beta}\bigg[\sqrt {4\beta \frac{\omega^2}{c^2}+(\gamma^2-4
\alpha \beta )q_x^2}-\gamma q_x\bigg], \label{qz}
\end{equation}
the choice of sign of $q_z$ ensures that light power propagates
away from the surface to the $+z$ direction. Due to the
anisotropy, the transmitted wave components may refract at
slightly different angles. The values of refracted wave vector can
be found by using the boundary condition and hyperbolic dispersion
relation.

We now determine the angle of phase refraction. The incident angle
of light in free space is $\theta_I =\tan^{-1}(k_{x}/k_{z})$ and
the refraction angle of the transmitted wave vector in AMM can be
found by $\theta_T= \tan^{-1}(q_{x}/q_{z})$. From the boundary
condition at the interface $z=0$, the tangential components of the
wave vectors must be continuous, i.e., $q_{x}=k_{x}$. Thus the
refracted angle of wave vector in the AMM can be easily obtained.

It is well known that the group velocity in anisotropic media
differs from the direction of its wave vector. The group velocity
is a very important physical quantity because it identifies the
speed of the maximum intensity of wave packet. The group velocity
in anisotropic media can be defined as ~\cite{Chen1983,Kong1990}
\begin{equation}
{\bf v}_{g}=\nabla_{\bf q}\omega({\bf
q})=\frac{\partial\omega}{\partial q_x}{\bf e}_x+
\frac{\partial\omega}{\partial q_z}{\bf e}_z, \label{vg}
\end{equation}
where $\nabla_{\bf q}$ denotes the gradient of $\omega({\bf q})$
in the wave vector space, ${\bf e}_x$ and ${\bf e}_z$ are unit
cartesian vectors. Because of the anisotropy, ${\bf
v}_g=\nabla_{\bf q}\omega({\bf q})$ is not necessarily parallel to
the wave vector ${\bf q}$.

The group velocity specifies the velocity of the peak of the wave
group, and is not necessarily parallel to the wave vector. When
the interface is aligned an angle with the optical axes of the
AMM, the hyperbolic dispersion relations will exhibit some
interesting effects. As shown in Fig.~\ref{Fig1}, the magnitude of
${\bf q}$ varies as a function of its direction. When the
refractive wave vector ${\bf q}$ is approximately parallel to the
asymptotic line of hyperbola(dash-dot-doted line), ${\bf q}$ can
be very large.  The wave front travels in the AMM with the
velocity of $v_p=\omega/q$, so the ultra-slow phase velocity can
be expected in the medium with hyperbolic dispersion relation. The
waves propagate in different directions or in the AMM, $\Delta
{\bf q}$ may not be parallel to ${\bf q}$. Therefore, the group
velocity is generally not parallel to the phase velocity. If ${\bf
q}$ propagates in some special direction, $\Delta q \rightarrow 0$
and the superluminal group velocity can be deduced. It should be
mentioned that the interesting properties never exist in the
medium with circle or ellipse dispersion relation.

\section{Superluminal group propagation}\label{sec3}
The negative refraction index imposes that the SRR presents
frequency dispersion~\cite{Smith2000,Shelby2001}. This, in turn,
conveys that actual anisotropic SRR is highly lossy. So the
superluminal group velocity is not easily experimentally verified
in SRR metamaterial. However, an extremely promising material,
such as the TGS crystal, satisfies the conditions
$\varepsilon_x>0, \varepsilon_z<0$. In principle, at far infrared
frequencies due to polarization dependence of certain phonon
resonances, the material with suitably low absorption
exist~\cite{Gerbaux1998,Dumelow2005}. Thus the TGS slab can be
prepared at low temperature and the superluminal group velocity
can be measured experimentally.

\begin{figure}
\includegraphics[width=10cm]{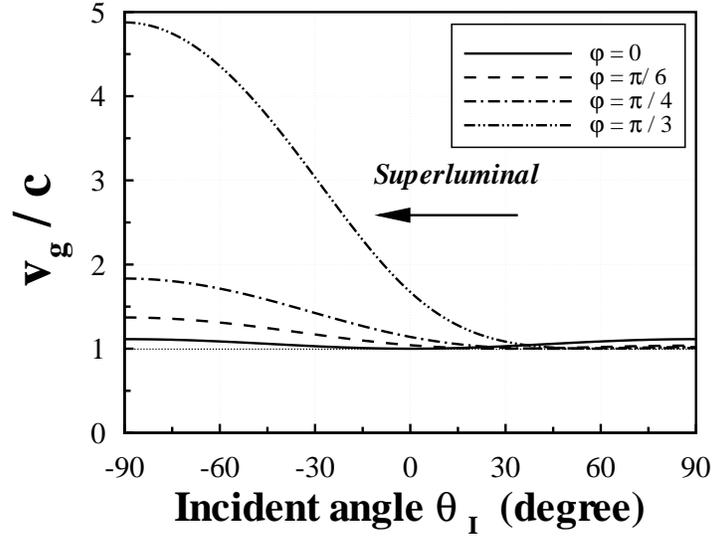}
\caption{\label{Fig2} The superluminal group velocity in an AMM
with different optical axial angles of $\varphi=0$ (solid),
$\pi/6$ (dashed), $\pi/4$ (dash-doted) and $\pi/3$
(dash-dot-doted). The parameters of AMM are choose as
$\varepsilon_x=1$, $\varepsilon_z=-5$ and $\mu_y=1$ (TGS
crystal).}
\end{figure}

Based on the dispersion relation of Eq.~(\ref{D2}), the group
velocity can be derived as
\begin{equation}
v_{g}=\frac{c(2 \alpha q_x +\gamma q_z){\bf e}_x+c( \gamma q_x +2
\beta q_z){\bf e}_z}{2\sqrt{\alpha q_x^2+\gamma q_x q_z+ \beta
q_z^2}} \bigg|_{\bar{\bf q}}. \label{vg}
\end{equation}
If the width of the distribution  ${\bf q}({\bf k}) $ is small
compared to the range of ${\bf k}$ over which ${\bf k}_0$ varies
significantly, we can simply evaluate $\bar{\bf q}({\bf k})={\bf
q}({\bf k}_0)$ to obtain a good approximation. In Fig.~\ref{Fig2},
the group velocity in the AMM is plotted. It should be noted that
the superluminal group velocity in the AMM is induced by the
hyperbolic dispersion relation. The superluminal group velocity
does not means the propagation of energy will be superluminal.
Using the new definition of electric and magnetic
energy~\cite{Cui2004a,Cui2004b}, the energy velocity can not be
superluminal and the conservation of energy is not violated.

Note that a wave group can be formed from planes with different
frequencies $\omega$ or from plane waves with different $\bf k$
vectors~\cite{Kong1990}. To obtain a better physical picture of
superluminal group velocity in AMM, a modulated Gaussian beam of
finite width can be constructed. The field intensity distribution
in free space is obtained by the Fourier integral and angular
spectrum representation~\cite{Goodman1988}. Following the method
outlined by Lu {\it et al.}~\cite{Lu2004}, let us consider a
modulated beam is incident from free space
\begin{equation}
H _I(x, z) = \int_{-\infty}^{+\infty}d k_\perp f( k_\perp)
\exp[i({\bf k}_0+{\bf k}_\perp) \cdot {\bf r}- i \omega_0  t],
\label{ft}
\end{equation}
where ${\bf k}_\perp$ is perpendicular to ${\bf k}_0$ and
$\omega_0=c k_0$. A general incident wave vector is written as
${\bf k}={\bf k}_0+{\bf k}_\perp $. we assume its Gaussian weight
is
\begin{equation}
f (k_\perp) = \frac{a}{\sqrt{\pi}} \exp [- a^2 k_\perp ],
\label{ft}
\end{equation}
where $a$ is the spatial extent of the incident beam. We want the
modulated Gaussian beam to be aligned with the incident direction
defined by the vector ${\bf k}_0=k_0 \cos \theta_I {\bf e}_x+ k_0
\sin \theta_I {\bf e}_z$, which makes an angle $\theta_I$ with the
surface normal.

\begin{figure}
\includegraphics[width=10cm]{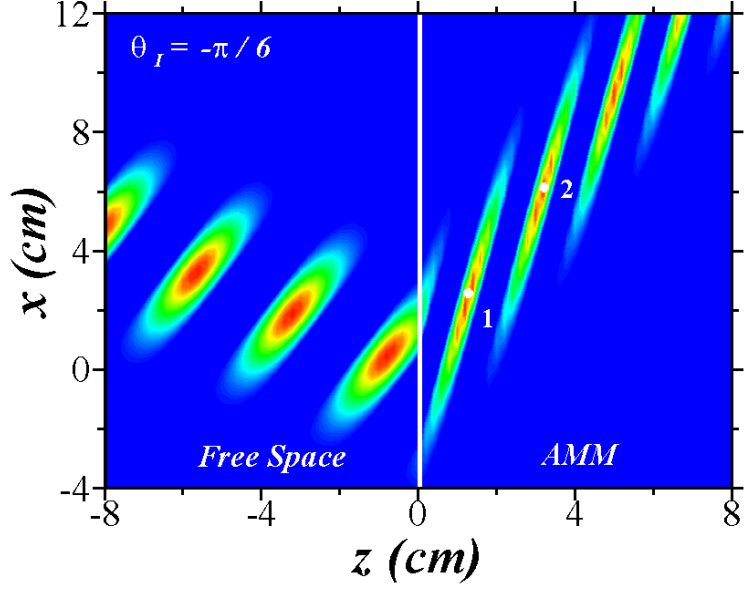}
\caption{\label{Fig3} The modulated Gaussian beam incident from
free space to AMM. The center wave number is $k_0=\sqrt{5}$ with
different incident angle $\theta_I=-\pi/6$. The optical axis angle
$\varphi=\pi/4$ and spatial extent of the incident wave packet is
$a=1$. The peak of localized wave packet travels from point $1$ to
point $2$ during a cycle $\Delta t$ with superluminal group
velocity. The anisotropic parameters are same as those used in
Fig.~\ref{Fig2}.}
\end{figure}

Due to the anisotropy, the transmitted wave components may refract
at slightly different angles. When the beam enters the AMM, it
will no longer maintain Gaussian, but becomes a localized wave
packet. Matching the boundary conditions for each $k$ component at
$z=0$ gives the complex field in the form
\begin{equation}
H_T(x, z) = \int_{-\infty}^{+\infty}d k_\perp f( k_\perp )T ({\bf
k}) \exp[i({\bf q}\cdot {\bf r}-\omega_0 t)].\label{er}
\end{equation}
where $T ({\bf k})$ is the transmission coefficient of the wave
packet. The transmission coefficient can obtain a good
approximation to simply evaluate this quantity at ${\bf k}_0$.
Then the transmission coefficient of the wave packet is simply
given by $T({\bf k}_0)$. The normal component of refracted wave
vector $q_z$ can be expanded in a Taylor series to first order in
${\bf k}_0 $ to obtain a better approximation
\begin{equation}
q_z({\bf k}) =q_z({\bf k}_0 )+({\bf k}-{\bf k}_0 )\cdot
\frac{\partial q_z({\bf k})}{\partial {\bf k}} \bigg|_{{\bf
k}_0}.\label{te}
\end{equation}
The distribution of the transmission field in the AMM can be
derived from Eq.~(\ref{er}) under the above approximation.

A close look at the localized wave packet shows that the
superluminal propagation of the peak is induced by the hyperbolic
dispersion relation in the AMM. Fig.~\ref{Fig3} shows a closes
view of the field intensity distribution of the wave packet
propagating from the free space into the AMM. In a circle $\Delta
t=2 \pi/ \omega_0$ of the modulated Gaussian beam, the peak of
localized wave packet travels from point $1$ to point $2$. In
Fig.~\ref{Fig3} we set the center wave vector with a incident
angles $\theta_I=-\pi/6$. We mark on each position of the peak of
wave packet at each of the two times. In the $\Delta t=0.094 ns$,
the peak of wave packet moves from $(1.284,~2.571)$ to
$(3.144,~6.147)$. This propagating velocity corresponds to $0.427
m/ns$, which is almost exactly the analytical group velocity of
$1.424 c$ in Eq.~(\ref{vg}), and the superluminal group velocity
is demonstrated theoretically.

\section{Discussion and conclusion}\label{sec4}
It should be noted that the superluminal group velocity has
completely different origin from those described in isotropic
LHM~\cite{Ziolkowski2001,Gupta2004,Woodley2004} or ultracold gas
of atoms~\cite{Chiao1993, Wang2000}, where the superluminality is
caused by the frequency dispersion of the medium. In the case
discussed here, the superluminal group velocity in the medium is
induced by the hyperbolic dispersion relation in the AMM. As far
as we know, this kind of superluminal group velocity has not been
recognized before. It should be mentioned, however, that the shape
of localized wave packet is distorted once the modulated Gaussian
beam is incident into the AMM. Our results do not violate
causality or special relativity, because the group velocity is the
velocity of the peak of the wave packet, which does not carry
information. It is shown that the TGS crystal slab can be designed
and the superluminal group velocity can be measured
experimentally.

\begin{acknowledgements}
H. Luo is sincerely grateful to Professor Y. Feng for many
fruitful discussions. This work was supported by projects of the
National Natural Science Foundation of China (No. 10125521, No.
10535010 and  No. 60278013), the Fok Yin Tung High Education
Foundation of the Ministry of Education of China (No. 81058).
\end{acknowledgements}

\end{document}